\documentclass[conference]{IEEEtran}
\IEEEoverridecommandlockouts
\usepackage{cite}
\usepackage{amsmath,amssymb,amsfonts}
\usepackage{algorithmic}
\usepackage{graphicx}
\usepackage{textcomp}
\usepackage[table]{xcolor}
\usepackage{booktabs}
\usepackage{url}
\usepackage{multirow}
\usepackage{array}
\usepackage{tikz}
\usetikzlibrary{arrows.meta,positioning,calc}
\usepackage{pgfplots}
\pgfplotsset{compat=1.16}
\usetikzlibrary{shapes.geometric,fit,backgrounds,patterns}
\usepackage{tabularx}
\usepackage{array}
\usepackage{comment}
\usepackage{listings}

\usepackage[moderate,paragraphs=normal,tracking=normal]{savetrees}

\definecolor{lstkw}{HTML}{1F4E79}
\definecolor{lstcm}{HTML}{6A8759}
\lstdefinestyle{ptcode}{%
  basicstyle=\ttfamily\scriptsize,
  language=Python,
  keywordstyle=\color{lstkw}\bfseries,
  commentstyle=\color{lstcm}\itshape,
  showstringspaces=false,
  columns=fullflexible,
  keepspaces=true,
  breaklines=true,
  xleftmargin=5pt, xrightmargin=2pt,
  aboveskip=2pt, belowskip=2pt
}
\definecolor{tblhead}{HTML}{CFDDEB}   
\definecolor{tblband}{HTML}{EFF4F9}   
\newcommand{\hrow}{\rowcolor{tblhead}}
\definecolor{sevCat}{HTML}{B2182B}
\definecolor{sevAbo}{HTML}{EF8A62}
\definecolor{sevSil}{HTML}{F6C445}
\definecolor{sevHin}{HTML}{6BAED6}
\definecolor{sevNo}{HTML}{74C476}
\definecolor{boxCode}{HTML}{ECEFF1}
\definecolor{boxLLM}{HTML}{D6EAF8}
\definecolor{boxRule}{HTML}{D5F5E3}
\definecolor{boxFilt}{HTML}{FCF3CF}
\definecolor{boxTest}{HTML}{E5E9F0}
\definecolor{boxOrac}{HTML}{EBDEF0}
\definecolor{boxCrash}{HTML}{FADBD8}
\def\BibTeX{{\rm B\kern-.05em{\sc i\kern-.025em b}\kern-.08em
    T\kern-.1667em\lower.7ex\hbox{E}\kern-.125emX}}

\usepackage[most]{tcolorbox}

\tcbset{
  findingbase/.style={
    enhanced,
    breakable,
    colback=gray!5,
    colframe=black,
    fonttitle=\bfseries,
    boxrule=0.8pt,
    arc=2mm,
    left=2mm,
    right=2mm,
    top=1mm,
    bottom=1mm
  }
}

\newtcolorbox{finding}[2][]{%
  findingbase,
  title={#2},
  #1
}

\begin{document}

\title{Beyond Fixed Fault Models: Comparing LLM-Based and Rule-Based Fault Injection in OpenStack}

\author{%
\IEEEauthorblockN{%
Giuseppe De Rosa\IEEEauthorrefmark{1}\IEEEauthorrefmark{2},
Pietro Liguori\IEEEauthorrefmark{1},
Domenico Cotroneo\IEEEauthorrefmark{3}}

\IEEEauthorblockA{\IEEEauthorrefmark{1}
\textit{University of Naples Federico II}, Naples, Italy}
\IEEEauthorblockA{\IEEEauthorrefmark{2}
\textit{IMT School for Advanced Studies Lucca}, Lucca, Italy}
\IEEEauthorblockA{\IEEEauthorrefmark{3}
\textit{University of North Carolina at Charlotte}, Charlotte, NC, USA}

\IEEEauthorblockA{%
\{giuseppe.derosa20, pietro.liguori\}@unina.it,
d.cotroneo@charlotte.edu}
}

\maketitle

\begin{abstract}
Software Fault Injection (SFI) supports testing of cloud systems by introducing software defects and observing their manifestation. Rule-based injectors such as ProFIPy provide controlled and reproducible source-level mutations but require fault patterns to be encoded manually. Large Language Models (LLMs) offer a data-driven alternative by generating context-dependent software faults. We compare two code LLMs, Qwen2.5-Coder and DeepSeek-Coder, with ProFIPy in OpenStack's Nova and Cinder services. On shared injection targets, activation and observable-failure rates are comparable, but operational profiles differ: LLM-generated faults produce more Catastrophic outcomes on Nova, whereas ProFIPy produces more Silent and Multi-component effects. The sampled LLM outputs also differ in how they manifest failure, while showing greater agreement in their propagation scope. These findings show that LLM-based fault injection extends the behavioral coverage of fixed fault models without establishing general superiority, and that practical adoption still requires controlled generation, runtime validation, system-level oracles, and reproducible experimental provenance.
\end{abstract}


\section{Introduction}

Cloud platforms must remain available and preserve a consistent state despite software defects, partial failures, and complex interactions among distributed services. This is particularly challenging for Infrastructure-as-a-Service platforms such as OpenStack~\cite{openstack}, where a single operation may involve APIs, message queues, databases, hypervisors, storage backends, and asynchronous workflows. A defect may remain local, propagate across services, or produce an apparently successful response while leaving the infrastructure in an inconsistent state.

Software Fault Injection (SFI) evaluates robustness by deliberately introducing defects and observing whether the resulting errors are activated, detected, tolerated, contained, or exposed as failures~\cite{natella2013representativeness,natella2016survey}. Source-level SFI targets \emph{residual faults}: implementation defects that escape testing and code review, remain latent in deployed software, and surface only under specific execution conditions~\cite{derosa2026break, cotroneo2026escape}.

Their effects are not always immediately visible. For example, an OpenStack volume-attachment request may return a successful API response even though the volume is unusable. Previous experiments detected such non-fail-stop failures only through independent state assertions~\cite{howbad}. Assessing cloud robustness, therefore, requires both representative faults and complementary oracles able to capture client-visible errors, internal failures, and inconsistent system state.

The value of an SFI campaign largely depends on its fault model. Arbitrary syntactic mutations may not reflect defects made by developers and may lead to misleading conclusions~\cite{natella2013representativeness}. Rule-based injectors address this problem through catalogs derived from empirical bug studies and defect taxonomies~\cite{duraes2006emulation,chillarege1992odc}. They encode recurring patterns, such as omitted calls, incorrect parameters, wrong return values, and missing control flow, as transformations over the program's abstract syntax tree.

ProFIPy applies this approach to OpenStack source code~\cite{howbad,openstackfi,profipy}. Its catalog provides valid, reproducible, and precisely controlled mutations. However, the explored fault space is bounded by manually designed operators, and context-dependent defects that do not match an existing template may remain unexplored.

Code Large Language Models (LLMs)~\cite{qwen,deepseek} offer a generative alternative. After learning from historical residual bugs, they can use the surrounding function context to generate implementation-specific faults rather than instantiate fixed transformations. This may reduce operator-authoring effort and broaden the explored fault space, but generated outputs may be invalid, equivalent, off-target, or incompatible with the target runtime~\cite{qiqo}. They therefore require explicit validation.

Existing work on learned mutation and LLM-based fault generation has mainly examined code-level properties, such as syntactic validity, similarity to historical defects, and mutation score~\cite{llmorpheus, smart, tufano2019mutate}. These properties do not establish whether an accepted fault exposes meaningful resilience weaknesses in a deployed distributed system. It remains unclear whether generative fault models reveal operational behaviors beyond those captured by fixed catalogs or primarily produce immediate, easily observable crashes.

We investigate this question through a controlled comparison of generative and catalog-based source-level fault injection in OpenStack. Rather than asking whether an LLM can produce a syntactically different function, we examine whether its accepted faults extend the \emph{operational behavior space} observed under a common deployment, workload, and set of failure oracles. Here, \emph{extension} denotes additional behaviors observed in the evaluated campaign.

We study the Nova compute and Cinder block-storage services of OpenStack Pike. Qwen2.5-Coder and DeepSeek-Coder are fine-tuned on PyResBugs~\cite{pyresbugs} and compared with ProFIPy using the same testbed, workload, coverage measurements, and failure oracles~\cite{profipy}. Since validation produces different accepted target sets, we report both campaign-level results and a location-controlled analysis over functions shared by all three injectors.

We address the following research questions:

\begin{itemize}
\item \textbf{RQ1 (Failure behavior).} How do generative and rule-based fault models differ in the failures they expose?
\item \textbf{RQ2 (Failure propagation).} How do generative and rule-based faults differ in their propagation across system components?
\item \textbf{RQ3 (Generative diversity).} How much does the choice of the generative model affect the operational behavior of injected faults?
\end{itemize}

Our results show that the two approaches expose complementary behaviors. On shared Nova targets, accepted LLM-generated faults cause more service-level crashes, whereas ProFIPy produces more Silent failures, in which an operation appears successful despite an inconsistent resource state. Throughout the campaign, ProFIPy faults propagate more frequently among components, whereas LLM-generated effects are predominantly local or result in immediate service disruption. Qwen and DeepSeek also produce different failure outcomes at the same locations, although their system-level impact is more strongly shaped by the target service and injection point. Overall, generative fault models broaden contextual exploration while complementing the control of fixed operator catalogs.
\section{Background \& Related Work}
\label{sec:related}

\subsection{Software Fault Injection}

Software-implemented fault injection operates at levels from processor state to source code and service interactions~\cite{hsueh1997fault,natella2016survey}. Low-level corruption commonly models transient faults~\cite{cinque2025cosmos}, whereas source transformations emulate persistent implementation defects. We use the latter and observe their effects through an end-to-end workload.
Source-level injection resembles mutation testing because both create modified program versions, but their goals differ~\cite{jia2011mutation, natella2016survey}. Mutation testing assesses a test suite's ability to kill mutants, whereas dependability studies examine activation, detection, containment, reporting, and propagation. A useful testing mutant is therefore not necessarily representative of a residual production fault.

Representativeness depends on the transformation, location, activation conditions, context, and failure behavior~\cite{natella2013representativeness,costa2015faultloads,just2014mutants}. ODC classifies defect semantics~\cite{chillarege1992odc}, while G-SWFIT derives operators such as omitted calls, wrong parameters, and missing control flow from field defects~\cite{duraes2006emulation}. Catalogs offer control and repeatability but may miss context-dependent faults; ProFIPy makes them programmable~\cite{profipy}.

Other approaches include distributed systems-based, that instead inject runtime failures: FATE and DESTINI target storage and I/O~\cite{fate}; Molly uses data lineage~\cite{ldfi}; Filibuster explores service calls~\cite{filibuster}; and Legolas selects locations using program states~\cite{legolas}. They primarily model exceptions, delays, failed calls, message loss, and storage faults rather than implementation defects. They complement the source-level faults examined here.

\subsection{Learned and Generative Fault Models}

Data-driven methods learn transformations from historical defects. Tufano et al.~\cite{tufano2019mutate} reversed a model trained on bug fixes to generate defective code.
Neural Fault Injection uses natural-language prompts~\cite{neuralfi}; PyResBugs pairs residual Python bugs with fixes and descriptions~\cite{pyresbugs}; and LLMorpheus applies LLMs to JavaScript mutation testing~\cite{llmorpheus}.

These methods extend fixed catalogs but can produce invalid, equivalent, runtime-incompatible, or unrepresentative changes. Existing work, therefore, emphasizes syntax, similarity to historical changes, mutant survival, or mutation score. Those metrics characterize generation and testing utility, not the operational consequences of activating a generated fault in a deployed system.

This distinction separates our study from LLM mutation testing. Our goal is to deploy each accepted mutation in a multi-service cloud and observe its effects across components and user-visible state, rather than evaluate the system's test suite.

To our knowledge, no study compares catalog-driven and generative source-level injection in one deployed distributed system under a common workload, deployment, and oracle configuration. We fill this gap by deploying accepted mutations from models fine-tuned on residual Python bugs and testing whether they extend the observed severity, visibility, and propagation space of a fixed catalog.

\section{Methodology}
\label{sec:design}

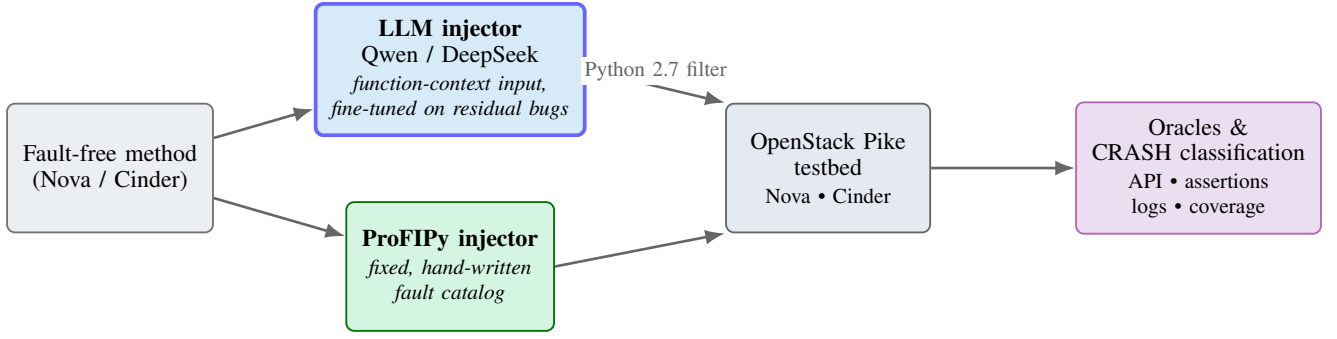
\begin{figure*}[t]
\centering
\begin{tikzpicture}[
  >=Latex, font=\small,
  box/.style={draw, rounded corners=3pt, align=center, inner sep=6pt,
              minimum height=17mm, minimum width=27mm, line width=0.7pt},
  code/.style={box, fill=boxCode, draw=black!55},
  llm/.style={box, fill=boxLLM, draw=blue!60, line width=1.4pt},
  rule/.style={box, fill=boxRule, draw=green!45!black},
  test/.style={box, fill=boxTest, draw=black!60},
  outb/.style={box, fill=boxOrac, draw=violet!60},
  flow/.style={->, line width=1pt, black!60},
  elab/.style={font=\footnotesize, text=black!60, fill=white, inner sep=1.5pt}
]
\node[code] (m)   at (0,0)      {Fault-free method\\(Nova / Cinder)};
\node[llm]  (llm) at (4.5,1.3)  {\textbf{LLM injector}\\Qwen / DeepSeek\\[1pt]{\footnotesize\itshape function-context input,}\\{\footnotesize\itshape fine-tuned on residual bugs}};
\node[rule] (pf)  at (4.5,-1.3) {\textbf{ProFIPy injector}\\[1pt]{\footnotesize\itshape fixed, hand-written}\\{\footnotesize\itshape fault catalog}};
\node[test] (t)   at (9.5,0)    {OpenStack Pike\\testbed\\[1pt]{\footnotesize Nova \textbullet\ Cinder}};
\node[outb] (o)   at (14.4,0)   {Oracles \&\\CRASH classification\\[1pt]{\footnotesize API \textbullet\ assertions}\\{\footnotesize logs \textbullet\ coverage}};
\draw[flow] (m) -- (llm);
\draw[flow] (m) -- (pf);
\draw[flow] (llm.east) -- node[above, elab] {Python~2.7 filter} (t.north west);
\draw[flow] (pf.east)  -- (t.south west);
\draw[flow] (t) -- (o);
\end{tikzpicture}
\caption{Experimental harness. Qwen and DeepSeek generate candidate mutations, whereas ProFIPy applies a controlled operator. Accepted Python~2.7-compatible mutations run on the same testbed and are classified from four oracles.}
\label{fig:pipeline}
\end{figure*}

This section details the methodology (Figure~\ref{fig:pipeline}) underlying our comparison. We evaluate the selected, fine-tuned LLMs and ProFIPy over a shared pool of target functions, using the same testbed adopted to evaluate ProFIPy~\cite{profipy} as workload, and four failure oracles to classify the observed outcomes.

\subsection{Dataset}
\label{subsec:dataset}

PyResBugs~\cite{pyresbugs} is the dataset used to fine-tune the evaluated LLMs and generate the faults. It contains 5{,}007 hand-validated residual bugs mined from dozens of open-source Python projects. Each record pairs a fault-free function with the faulty version that developers left in production, allowing models to learn real rather than synthetic defects. Its G-SWFIT~\cite{duraes2006emulation} and ODC~\cite{chillarege1992odc} labels cover ProFIPy's fault families: wrong parameters, omitted calls, incorrect returns, and missing handlers.

We train only on code pairs to teach the models how to generate software faults without any further natural-language instructions. The model, therefore, learns a direct clean-to-faulty transformation. We verified that neither the target functions nor their bug-fixing commits occur in the fine-tuning set.

The leakage check compares the normalized target source and associated commit identifiers with every fine-tuning record. A shared project or function name alone is not considered leakage because common names can occur in unrelated modules; a match requires the same target code or fixing commit.

\subsection{Fault Generation}

We fine-tune Qwen2.5-Coder-32B-Instruct~\cite{qwen} and DeepSeek-Coder-33B~\cite{deepseek} with QLoRA~\cite{qlora,lora}. Both are code-specialized models of comparable size, so the comparison does not conflate generation strategy with a general-purpose model. Both use 4-bit NF4 weights, double quantization, bfloat16 computation, rank $r=16$, LoRA $\alpha=32$, and dropout $0.05$ on the query, key, value, output, gate, up, and down projections, trained for 2--3 epochs with early stopping on validation loss, a batch size of 1 with 8-step accumulation, a learning rate of $2\times10^{-4}$, 4{,}096-token sequences, 8-bit paged AdamW, a cosine schedule with 3\% warm-up, and seed 42.

For each generation, each target code block receives the identical instruction \emph{``Inject a single and realistic software fault into the function,''} with a system message requiring syntactically valid Python, a changed body, and only the faulty function in one code block without explanation; keeping the prompt fixed across targets avoids introducing variation from prompt wording. Sampling uses nucleus sampling with temperature $0.4$, top-$p=0.95$, up to 1{,}024 new tokens, and up to three attempts per target.

Each candidate then passes through two validation stages before being accepted. The first checks that the output is not empty or whitespace-equivalent and that its AST contains a function definition. This distinguishes a usable function from prose, an empty response, or an unchanged copy of the input. The second stage filters for Python~2.7 compatibility, since targets must run under OpenStack Pike. Neither stage verifies semantic equivalence, signature preservation, or whether the rewrite introduces exactly one independent semantic edit; they establish only that a candidate compiles and is deployable, leaving operational behavior to be determined later by the workload and oracles. The first candidate to pass both stages enters the campaign, and targets that fail all three attempts are dropped rather than resampled indefinitely.


\subsection{Measurement and Analysis}

OpenStack studies report inconsistent state, delayed reporting, non-fail-stop behavior, and cross-service propagation~\cite{openstackresilience,howbad}. A client-visible API response alone can therefore misrepresent the final resource state, while a local service log may miss effects that travel through the control plane. These findings motivate four oracles: API, state assertions, logs, and coverage.
 
The API oracle records HTTP 4xx/5xx responses; in-guest assertions test functional state, including volume writes and reads; Syslog entries at WARNING or higher reveal internal errors; and \texttt{coverage.py}~\cite{coveragepy} records whether mutated code executes. The oracles are complementary: API responses capture client-visible failures but cannot reveal an unusable resource reported as created. Assertions expose internal silent failures, logs capture errors that are not propagated to clients, and coverage distinguishes activated faults.

We used these oracles in sequence to identify the failure outcome. A fault may be dormant, i.e., the workload could never exercise it. Thus, coverage first determines whether the target executes (i.e., verifies the fault activation). If it does, the API, assertion, and log observations distinguish an absorbed fault from a manifest failure by examining, respectively, client-level and system-level states. 

Outcomes are classified using the Koopman CRASH scale~\cite{crash}: Catastrophic, Restart, Abort, Silent, and Hindering. We adapt these classes to cloud services by defining each in terms of its observable effect, since a distributed service can fail in ways a standalone program cannot (e.g., an operation that returns success while leaving persistent state inconsistent). Runs where the fault is never triggered (dormant) and runs where it is triggered but produces no oracle-visible failure (activated) both fall under No-failure; we report them separately in the results to distinguish untested code from code that tolerated the fault.

\begin{table}[t]
\caption{CRASH modes, cloud manifestations, and detecting oracles. No Restart occurred.}
\label{tab:crashmap}
\centering
\footnotesize
\rowcolors{2}{tblband}{white}
\begin{tabular}{@{}lp{0.42\columnwidth}p{0.22\columnwidth}@{}}
\toprule
\hrow \textbf{Class} & \textbf{Cloud manifestation} & \textbf{Oracle} \\
\midrule
Catastrophic & Service daemon crashes at start-up; workload blocked & Log, API \\
Restart & Operation hangs and requires a restart (not observed) & Timeout \\
Abort & Operation fails immediately with an API error & API error \\
Silent & HTTP~2xx success, but the resulting resource is unusable or inconsistent & Assertion \\
Hindering & Failure is detected late; an assertion fails before a subsequent API error & Assertion, API \\
No failure & Target is dormant, or activation yields no oracle-visible failure & Coverage/none \\
\bottomrule
\end{tabular}
\end{table}

Classification is based on the end-to-end manifestation observed during the workload. Concretely: a crash of a service daemon at start-up is Catastrophic; an operation rejected immediately with an API error is Abort; an HTTP~2xx response followed by an invalid or inconsistent resource is Silent; and a failure caught by a late assertion, followed by a subsequent API error, is Hindering. Restart is reserved for hangs or timeouts requiring recovery; this class did not occur in our campaign. Table~\ref{tab:crashmap} summarizes each class, its cloud-specific manifestation, and the oracle used to detect it.

Finally, an independent dimension complements the CRASH scale. IMPACT records propagation: the number of components affected by the faulty run, other than the original component where the fault is injected. Total for a system-wide effect, Multi when an effect reaches another component, Local when an activated run remains confined to the injected component, and Dormant when the target is not executed. Local, therefore, includes contained runs with no observed failure. Keeping CRASH and IMPACT separate lets us differentiate severity and propagation.


\subsection{Experimental Design}

For comparability with ProFIPy, the testbed runs OpenStack Pike on CentOS~7 and Python~2.7. It comprises a Controller that hosts RabbitMQ, MariaDB, Keystone, Glance, and the Nova and Cinder control services; a Compute node runs \texttt{nova-compute} over KVM/QEMU. We inject into Nova compute and driver managers and Cinder's volume manager and driver, which implement the compute and block-storage operations exercised by the workload. These services also use RabbitMQ and shared control-plane workflows, making them suitable for observing propagation.

All suites run the workload~\cite{ofie} used in the ProFIPy study~\cite{profipy, howbad} for comparability. Through REST APIs, it uploads CirrOS, provisions a private network, router, and floating IP, boots a VM, and creates and attaches a Cinder volume. It checks SSH reachability, reads and writes the attached block device, verifies a hard reboot, and deletes all resources.

A fault is \emph{activated} when the workload reaches its injection point and \emph{dormant} otherwise; coverage records this relative to a fault-free run. Activation is not guaranteed: some faults may introduce triggering conditions that the workload cannot meet.

Also, we highlight that activation does not imply a failure. OpenStack may catch an injected error, roll back an operation, or return a controlled response without corrupting the observed state. An \emph{activated, no-observed-failure} run executes but leaves no trace in the API, assertion, or log oracles; it may be absorbed, irrelevant to this workload, or semantically equivalent. The experiment does not distinguish these causes. A \emph{manifest} failure executes and is flagged by at least one oracle, after which it is classified on the CRASH severity scale.

The campaign has 122 one-target scenarios: 66 on Nova and 56 on Cinder (Table~\ref{tab:scenarios}). Each deploys one mutated function in one source file. A fault-free checkout is performed after every injection to avoid environmental drift or resources left by earlier runs; the idea is to clean the environment before continuing, so carryover is not attributed to the next mutation.

\begin{table}[t]
\caption{Scenario inventory; one mutated target per scenario.}
\label{tab:scenarios}
\centering
\footnotesize
\rowcolors{2}{tblband}{white}
\begin{tabular}{@{}lcccc@{}}
\toprule
\hrow \textbf{Service} & \textbf{ProFIPy} & \textbf{DeepSeek} & \textbf{Qwen} & \textbf{Total} \\
\midrule
Nova   & 23 & 21 & 22 & 66 \\
Cinder & 20 & 17 & 19 & 56 \\
\midrule
Total  & 43 & 38 & 41 & 122 \\
\bottomrule
\end{tabular}
\end{table}

\section{Results}
\label{sec:results}

We compare the three injectors along three dimensions: the activation and
failure behavior of their faults, the propagation of their effects across
components, and the variability between the two generative models.

\begin{table}[t]
\caption{CRASH-mode percentages for the full suites (sizes in parentheses; PF: ProFIPy, DS: DeepSeek, Qw: Qwen). No failure combines dormant and activated/no-observed runs.}
\label{tab:crash}
\centering
\footnotesize
\rowcolors{4}{tblband}{white}
\setlength{\tabcolsep}{4.5pt}
\begin{tabular}{@{}lccc@{\hspace{7pt}}ccc@{}}
\toprule
\hrow  & \multicolumn{3}{c}{\textbf{Nova}} & \multicolumn{3}{c}{\textbf{Cinder}} \\
\cmidrule(lr){2-4}\cmidrule(lr){5-7}
\hrow \textbf{Failure mode} & PF & DS & Qw & PF & DS & Qw \\
\hrow  & \scriptsize(23) & \scriptsize(21) & \scriptsize(22) & \scriptsize(20) & \scriptsize(17) & \scriptsize(19) \\
\midrule
Catastrophic & 4.3  & 28.6 & 22.7 & 0.0  & 0.0  & 0.0 \\
Abort        & 17.4 & 19.0 & 9.1  & 5.0  & 11.8 & 21.1 \\
Silent       & 30.4 & 9.5  & 18.2 & 0.0  & 0.0  & 5.3 \\
Hindering    & 17.4 & 4.8  & 18.2 & 20.0 & 23.5 & 26.3 \\
No failure   & 30.4 & 38.1 & 31.8 & 75.0 & 64.7 & 47.4 \\
\bottomrule
\end{tabular}
\end{table}

\subsection{RQ1: How do generative and rule-based fault models differ in the failures they expose?}

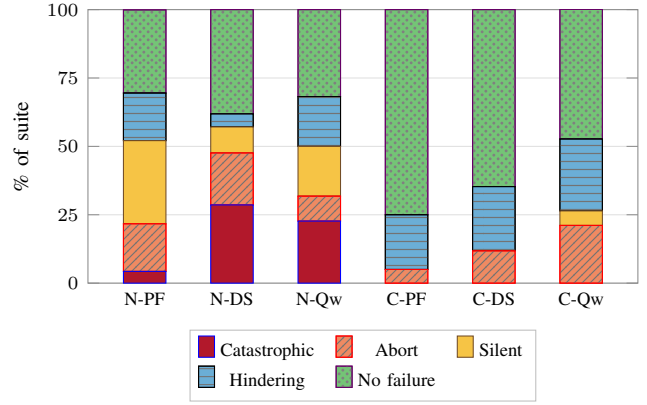
\begin{figure}[t]
\centering
\begin{tikzpicture}
\begin{axis}[
  ybar stacked, bar width=16pt,
  width=\columnwidth, height=5.2cm,
  ymin=0, ymax=100,
  ylabel={\footnotesize \% of suite}, ylabel near ticks,
  symbolic x coords={N-PF,N-DS,N-Qw,C-PF,C-DS,C-Qw},
  xtick=data, x tick label style={font=\scriptsize},
  ytick={0,25,50,75,100}, y tick label style={font=\scriptsize},
  ymajorgrids, grid style={draw=black!12, thin},
  axis line style={draw=black!45}, tick style={draw=black!45},
  every axis plot/.append style={draw=black!45, line width=0.4pt},
  legend style={font=\scriptsize, at={(0.5,-0.17)}, anchor=north, legend columns=3,
                draw=black!25, /tikz/every even column/.append style={column sep=6pt}},
  enlarge x limits=0.13,
]
\addplot+[fill=sevCat] coordinates {(N-PF,4.3)(N-DS,28.6)(N-Qw,22.7)(C-PF,0)(C-DS,0)(C-Qw,0)};
\addplot+[fill=sevAbo, postaction={pattern=north east lines, pattern color=black!55}] coordinates {(N-PF,17.4)(N-DS,19.0)(N-Qw,9.1)(C-PF,5.0)(C-DS,11.8)(C-Qw,21.1)};
\addplot+[fill=sevSil] coordinates {(N-PF,30.4)(N-DS,9.5)(N-Qw,18.2)(C-PF,0)(C-DS,0)(C-Qw,5.3)};
\addplot+[fill=sevHin, postaction={pattern=horizontal lines, pattern color=black!55}] coordinates {(N-PF,17.4)(N-DS,4.8)(N-Qw,18.2)(C-PF,20.0)(C-DS,23.5)(C-Qw,26.3)};
\addplot+[fill=sevNo, postaction={pattern=crosshatch dots, pattern color=black!50}] coordinates {(N-PF,30.4)(N-DS,38.1)(N-Qw,31.8)(C-PF,75.0)(C-DS,64.7)(C-Qw,47.4)};
\legend{Catastrophic, Abort, Silent, Hindering, No failure}
\end{axis}
\end{tikzpicture}
\caption{CRASH distributions for the full suites (N: Nova, C: Cinder; PF: ProFIPy, DS: DeepSeek, Qw: Qwen).}
\vspace{-4mm}
\label{fig:crashbars}
\end{figure}

Table~\ref{tab:crash} and Figure~\ref{fig:crashbars} summarize the CRASH outcomes produced by the three injectors. We first examine the full experimental suites and then compare only the functions successfully targeted by all three injectors. 

Across the full suites, accepted LLM-generated mutations activate in $77/79$ runs, compared with $39/43$ ProFIPy mutations. An observable failure occurs in $44/79$ LLM runs and $21/43$ ProFIPy runs. These aggregate values describe the behavior of the respective experimental suites, but part of the difference may result from the functions selected by each injector rather than from the fault-generation approach itself.

Comparing the 37 functions shared by all three injectors reduces this target-selection effect. Activation is similarly high for ProFIPy ($34/37$), DeepSeek ($36/37$), and Qwen ($37/37$), while observable failures occur in $19/37$, $18/37$, and $22/37$ runs, respectively. Thus, on common targets, the LLM-generated faults do not show a substantial activation advantage. However, the types of failures remain different. Among the 20 shared Nova functions, ProFIPy produces one Catastrophic failure, compared with six for DeepSeek and five for Qwen. Conversely, ProFIPy produces seven Silent failures, compared with two for DeepSeek and four for Qwen. A contrast in failure profiles, therefore, persists even when the target functions are held constant.

On Nova, the main difference concerns the severity and observability of the failures. DeepSeek and Qwen produce substantially more Catastrophic outcomes than ProFIPy: $28.6\%$ and $22.7\%$, respectively, compared with $4.3\%$. These faults often prevent a service from completing its start-up sequence, making the failure immediately visible. ProFIPy instead produces the largest proportion of Silent failures, at $30.4\%$, compared with $9.5\%$ for DeepSeek and $18.2\%$ for Qwen. In these cases, an operation reports success even though the resulting resource is invalid or unusable. Such failures are particularly relevant because the platform continues serving requests without exposing an explicit error.

Cinder exhibits a different failure profile. None of the injectors produces a Catastrophic outcome, and the distributions are dominated by Hindering failures ($20.0$--$26.3\%$) and runs with no observed failure ($47.4$--$75.0\%$). Cinder's exception-handling paths frequently contain the injected error, allowing the workload either to continue or to fail later through an explicit operation-level error. Consequently, the contrast between the injectors is less pronounced than on Nova.

The No failure category includes both faults that never activate and faults that activate without producing an observable effect. On Nova, the dormant faults account for four ProFIPy runs, one DeepSeek run, and one Qwen run; no dormant fault occurs on Cinder. After excluding these cases, the numbers of activated faults with no observed failure are 3, 7, and 6 on Nova, and 15, 11, and 9 on Cinder, for ProFIPy, DeepSeek, and Qwen, respectively. These activated faults may have been absorbed by exception handling, may affect behavior outside the exercised workload, or may be semantically equivalent to the original code.

Silent failures are distinct from these cases because they both activate and produce an observable inconsistency in the state. Although the LLM-generated suites contain fewer Silent failures overall, they expose failures tied to specific end-to-end workflows. For example, one Qwen mutation causes a volume-attachment request to return an HTTP 2xx response even though the attached volume is unusable. This inconsistency is detected only by the subsequent in-guest write/read assertion.

The differences between Nova and Cinder show that the manifestation of failure depends not only on the injected fault but also on the architecture of the target service. Nova exposes a broad severity range, including start-up crashes and silent resource corruption. Cinder more often contains exceptions or delays their effects, shifting the distribution toward Hindering outcomes and runs with no observed failure. Service architecture and exception handling, therefore, mediate the observable impact of all three fault models under the same workload and Oracle configuration.

Overall, accepted LLM-generated faults broaden the observed failure profile by producing more Catastrophic and workflow-specific outcomes. ProFIPy exposes more Silent failures and propagated effects. Moreover, the similar activation rates on shared functions show that the higher aggregate activation of the LLM suites cannot be interpreted as a generator-only effect.

\begin{finding}{}{}
\textbf{Key Takeaway:} \emph{On the same target functions, LLM-generated faults cause more visible crashes and workflow-specific failures, while ProFIPy reveals more silent corruptions. Differences in overall activation rates are partly due to the different functions targeted by each suite.}
\end{finding}{}

\subsection{RQ2: How do generative and rule-based faults differ in their propagation across system components?}

Table~\ref{tab:impact} shows that LLM-generated and rule-based faults differ not only in severity, but also in how far their effects spread. Accepted LLM-generated faults more often remain Local or produce a Total impact, whereas ProFIPy faults more often propagate to another component without causing a system-wide outage.

This pattern is clearer on Nova. DeepSeek and Qwen produce Total effects in $28.6\%$ and $22.7\%$ of the runs, respectively, compared with $4.3\%$ for ProFIPy. These outcomes largely correspond to start-up crashes that block the end-to-end workload. ProFIPy instead produces more Multi-component effects: $26.1\%$, compared with $9.5\%$ for DeepSeek and $13.6\%$ for Qwen.

Cinder shows the same tendency without Nova's large number of Total outcomes. ProFIPy produces Multi-component effects in $25.0\%$ of the runs, compared with $11.8\%$ for DeepSeek and $5.3\%$ for Qwen, while Local outcomes account for $75.0\%$, $88.2\%$, and $89.5\%$, respectively. Among manifest Cinder failures, all five ProFIPy cases affect another component. By contrast, four of six DeepSeek failures and eight of ten Qwen failures remain Local.

These differences are consistent with the structure of the injected mutations. ProFIPy typically applies small rule-based changes, often limited to one expression or statement. Such mutations can alter a value, condition, or return path while preserving execution, allowing the resulting inconsistent state to travel through REST calls, RabbitMQ messages, or shared control-plane workflows. LLM-generated mutations are often broader and may span multiple statements or modify control flow more substantially. They are therefore more likely to directly disrupt the injected service, leaving less opportunity for an intermediate state to propagate across components.

Severity and propagation also show a correlation, even though they measure different properties. Catastrophic failures often have Total impact because the loss of a service blocks the complete workload. Silent and Hindering failures are more often compatible with Multi-component propagation because execution continues long enough for an inconsistent state to reach another service. However, one dimension does not determine the other: a severe failure may remain Local, while a subtle mutation may propagate across multiple components. The Qwen Cinder case with Total impact but no daemon crash further shows that a system-wide effect does not necessarily imply a Catastrophic failure. CRASH, therefore, captures how the workload fails, whereas IMPACT captures how far the effect spreads.

\begin{finding}{}
\textbf{Key Takeaway:} \emph{ProFIPy's small rule-based mutations more often preserve execution and propagate across components, whereas broader LLM-generated mutations more often remain local or cause system-wide disruption. Severity and propagation are correlated, but capture distinct failure properties.}
\end{finding}

\begin{table}[t]
\caption{IMPACT percentages.}
\label{tab:impact}
\centering
\footnotesize
\rowcolors{3}{tblband}{white}
\setlength{\tabcolsep}{4.5pt}
\begin{tabular}{@{}lccc@{\hspace{7pt}}ccc@{}}
\toprule
\hrow  & \multicolumn{3}{c}{\textbf{Nova}} & \multicolumn{3}{c}{\textbf{Cinder}} \\
\cmidrule(lr){2-4}\cmidrule(lr){5-7}
\hrow \textbf{Scope} & PF & DS & Qw & PF & DS & Qw \\
\midrule
Total & 4.3  & 28.6 & 22.7 & 0.0  & 0.0  & 5.3 \\
Multi & 26.1 & 9.5  & 13.6 & 25.0 & 11.8 & 5.3 \\
Local & 52.2 & 57.1 & 59.1 & 75.0 & 88.2 & 89.5 \\
Dormant & 17.4 & 4.8  & 4.5  & 0.0  & 0.0  & 0.0 \\
\bottomrule
\end{tabular}
\end{table}

\subsection{RQ3: How much does the choice of the generative model affect the operational behavior of injected faults?}

DeepSeek and Qwen exhibit different failure profiles. On Nova, DeepSeek causes $6/21$ Catastrophic failures, with a $95\%$ Wilson interval of $14$--$50\%$, while Qwen causes $5/22$, with an interval of $10$--$43\%$. The wide and overlapping intervals indicate uncertainty. Descriptively, DeepSeek produces a larger Catastrophic share, whereas Qwen produces a more varied distribution and the largest Abort share on Cinder ($21.1\%$).

To separate model choice from injection location, we compare the faults generated at the 20 Nova and 17 Cinder locations shared by both models. The two models produce the same CRASH outcome in only $55\%$ of the shared Nova locations and $71\%$ of the shared Cinder locations. Thus, even at the same injection point, they often induce different failure manifestations.

Agreement is substantially higher for IMPACT: $80\%$ on Nova and $88\%$ on Cinder. This suggests that the injection location constrains how far a fault can spread more strongly than it constrains the exact way in which the workload fails. The model influences whether a location produces, for example, an Abort, Silent, or Catastrophic outcome, while the surrounding architecture often determines whether the effect remains Local or reaches other components.

The higher agreement on Cinder is consistent with its stronger error containment. Its exception-handling paths absorb or localize many mutations, reducing the observable differences between the models. At highly vulnerable locations, the surrounding code can dominate the outcome entirely: both models may produce a system-wide failure despite generating different mutations. Model-specific differences become more visible at locations where execution can continue, and the injected error can follow different control-flow or state-propagation paths.

Manual inspection also suggests different mutation styles. Some accepted DeepSeek faults make small interface-level changes, such as modifying parameters or call arguments, which can trigger failures at import, start-up, or call time. Qwen faults more often include larger body rewrites, additional branches, \texttt{try}/\texttt{except} blocks, or reimplemented logic, which can preserve execution and produce a wider range of later failures. These observations are qualitative because we did not classify every generated mutation using a systematic taxonomy.

Overall, similar validity and activation rates do not imply equivalent operational behavior. At the same target location, the two models can generate valid and activated faults that produce different CRASH outcomes. However, their higher agreement on IMPACT shows that the architecture and injection location remain major determinants of the resulting system-level effect.

\begin{finding}{}
\textbf{Key Takeaway:} \emph{Model choice affects how a fault manifests, and the two models show a different mutation strategy. DeepSeek and Qwen often differ in CRASH outcome while producing similar IMPACT scopes, primarily due to the injection location.}
\end{finding}

\section{Discussion}
\label{sec:discussion}

The evaluation suggests that LLMs help where effort is spent on fault-model construction. Rule-based injectors require fault patterns to be identified, formalized, and implemented as operators before a campaign can begin. An LLM can instead derive context-dependent mutations directly from the target function. This capability broadens the explorable fault space and makes application-specific mutations easier to generate.

However, context-specific mutation alone does not guarantee the representativeness of the injected fault. Fault realism must be assessed in relation to both the defect being modeled and the execution environment in which it is activated. The path, therefore, still requires control over mutation scope, evidence that the modified code executes, representative workloads, failure oracles, and sufficient provenance to reproduce the run. LLMs reduce operator-authoring effort, but they do not eliminate the engineering required to establish fault representativeness.

\subsection{From Fixed to Data-Driven Fault Models}

From our experience, an LLM should be treated as a generator of faults based on a fault model inferred from the training data. The inspected mutations show why this distinction matters. Generated changes may use the local context to alter exception handling, interfaces, branches, or several related statements. This flexibility can express defects that would require a dedicated rule-based operator, but it can also introduce multiple semantic changes or replace too much of the original implementation.

For fault-injection purposes, broader changes are not necessarily better. A mutation that immediately prevents a service from starting may expose a valid robustness weakness, but it exercises a different behavior from a small defect that preserves execution and allows an inconsistent state to propagate. A practical LLM-based injector should support explicit control over mutation granularity and campaign intent. For example, the generation process could request a single semantic change, avoid unrelated rewrites, or favor execution-preserving defects when the objective is to study error propagation. A realistic campaign should deliberately enable more and reproducible configurations to replicate fault realism and a thorough evaluation. 

Figure~\ref{fig:example} illustrates this. The LLM uses the surrounding exception-handling context to produce an architecture-specific change. The asynchronous request has already been acknowledged when the compute-side attachment fails, leaving a volume that appears attached but is unusable. ProFIPy instead applies an AST transformation that removes calls whose names match \texttt{volume}~\cite{profipy}. The operator is precise and reproducible, but it does not account for the semantic role of each matched call.

The practical advantage of the LLM is that the exception-handling rewrite does not need to be anticipated and implemented as a dedicated operator. Its limitation is that the generated change is not automatically atomic, representative, or reproducible.

\begin{figure}[t]
\begin{lstlisting}[style=ptcode]

# Fault-free

try:
    return self._attach_volume(context, instance, driver_bdm)
except Exception:
    with excutils.save_and_reraise_exception():
        bdm.destroy()

# LLM: exception-handler rewrite

- with excutils.save_and_reraise_exception():
-   bdm.destroy()

+ bdm.destroy()
+ raise

# ProFIPy: call-removal operator

- return self._attach_volume(context, instance, driver_bdm) 
+ return
  \end{lstlisting}
  \caption{Fault-free volume-attachment handler and source changes introduced by the LLM and ProFIPy.}
  \label{fig:example}
  \end{figure}

Hence, validation must also extend beyond parsing or compilation. It should distinguish at least four properties: (i) whether the mutation is syntactically valid and deployable; (ii) whether it preserves the intended mutation scope and target interface; (iii) whether the modified code is activated by the workload; (iv) whether activation produces an observable operational effect.

Table~\ref{tab:paradigms} summarizes these observations.

\subsection{Implications for Developers and Researchers}

For tool developers, the main implication is that generation should be embedded in a staged fault-injection pipeline. The model should propose a mutation, while deterministic components enforce structural constraints, check runtime compatibility, deploy the modified program, verify activation, and collect system-level observations. This division retains the flexibility of generative models without delegating the validity of the experiment entirely to the model.

Runtime feedback can further improve this process. Dormant mutations can be regenerated at covered locations, while changes that consistently cause immediate initialization failures can be balanced with prompts or constraints that favor execution-preserving behavior. Failure observations can also be used to select mutations according to campaign objectives, such as service recovery, silent-state corruption, exception containment, or cross-component propagation. The relevant optimization target is therefore not merely the probability of producing compilable code, but the probability of producing an activated and operationally informative fault.

For researchers, the results highlight the need to evaluate generative fault models at multiple levels. Code-level properties such as syntax, similarity to historical defects, or mutation score describe the generated artifact but not its behavior in a deployed system. Evaluations should additionally measure activation, failure visibility, severity, propagation, and final system state. Comparisons should control for injection location whenever possible because the target code and surrounding architecture can dominate the resulting behavior.

Generated faults also require stronger provenance than deterministic operators. Each run should retain the target location, prompt, generation attempts, validation decisions, deployment outcome, coverage, workload observations, and failure classifications. Without these records, differences due to generation variability cannot be distinguished from those due to the testbed or workload.

A hybrid lifecycle can combine exploration and reproducibility by using LLMs to discover application-specific mutations without requiring a dedicated operator for each defect, then minimizing, reviewing, and encoding recurrent or operationally relevant mutations as deterministic operators. In this way, the rule-based catalog becomes an evolving and reproducible core informed by generative exploration.

\begin{table}[t]
\caption{Key requirements for generative fault injection.}
\label{tab:paradigms}
\centering
\footnotesize
\rowcolors{2}{tblband}{white}
\begin{tabular}{@{}p{0.20\columnwidth}p{0.32\columnwidth}p{0.38\columnwidth}@{}}
\toprule
\hrow \textbf{Concern} & \textbf{Lesson} & \textbf{Recommended action} \\
\midrule

Fault model &
LLMs reduce manual operator authoring but may generate broad or compound changes &
Constrain mutation scope and preserve relevant interfaces \\

Representative-ness &
A plausible source change is not necessarily a realistic fault &
Ground generation in real bug data and review fault semantics \\

Operational validity &
Compilable mutations may be dormant or produce no observable effect &
Validate deployment, measure activation, and use end-to-end oracles \\

Reproducibility &
Generated faults depend on the model and sampling configuration &
Record generation settings, validation decisions, and source diffs \\

Campaign lifecycle &
Generated faults support exploration but offer limited repeatability &
Promote relevant generated patterns into deterministic operators \\

\bottomrule
\end{tabular}
\end{table}
\section{Threats to Validity}
\label{sec:threats}

\noindent\textbf{Construct validity.}
The CRASH scale required adaptation to distributed cloud services, and the distinction between Silent and Hindering outcomes depends on oracle coverage. Start-up crashes are also easier to detect than delayed state corruption. We mitigate these threats through explicit cloud-level definitions, complementary API, log, state, and coverage oracles, and a separate propagation classification to show their representativeness and failure outcomes.

\noindent\textbf{Internal validity.}
The Python~2.7 compatibility filter may bias the accepted LLM suites toward legacy-compatible faults. Campaign-level differences may also reflect different target locations; we therefore analyze the locations shared by all injectors, although they are not a random sample. Generation is stochastic, and only the first valid candidate within three attempts is evaluated, so differences between Qwen and DeepSeek may also reflect sampling. Given the limited sample sizes, comparisons are primarily descriptive; Wilson intervals quantify uncertainty around individual proportions but do not establish differences between injectors.

\noindent\textbf{External validity.}
The study covers two services, one OpenStack version, one workload, and 122 partly overlapping scenarios. Results may differ for other services, workloads, concurrency levels, platforms, or fault locations. We also evaluate only two code-specialized models, one training dataset, one generation configuration, and one rule-based injector. The findings should therefore be interpreted as evidence from the evaluated campaign rather than as universal superiority of either paradigm.

\section{Conclusion and Future Work}
\label{sec:conclusion}

This study shows that LLM-based fault injection can complement fixed-model SFI by generating context-dependent mutations without requiring a dedicated operator for each fault pattern. In the evaluated OpenStack campaign, LLM-generated faults broadened the observed failure space, while ProFIPy provided greater control, repeatability, and more execution-preserving and propagating failures. The contribution is therefore not generator-level superiority, but evidence that LLMs can reduce fault-model authoring effort and extend fixed catalogs.

Realistic generative fault injection still requires controlled mutation scope, deployment and activation validation, representative workloads, system-level oracles, and reproducible provenance. Future work will evaluate broader settings, incorporate runtime feedback into generation, and promote relevant generated faults into deterministic operators, toward an adaptive pipeline.

\bibliographystyle{IEEEtran}
\bibliography{biblio}

\end{document}